\documentclass[12pt]{iopart}

\usepackage{color}
\usepackage{iopams}  
\usepackage{graphicx}
\usepackage{url}
\usepackage{verbatim}
\begin{document}

\title[]{Method to estimate ISCO and ring-down
frequencies in binary systems and consequences for
gravitational wave data analysis}
\author{Chad Hanna$^{1}$, Miguel Megevand$^{1}$, Evan Ochsner$^{2}$, and Carlos Palenzuela$^{1,3}$}
\address{
$^{1}$ Department of Physics and Astronomy,
Louisiana State University
Baton Rouge, LA 70802, USA \\
$^{2}$ Department of Physics, University of Maryland,
College Park, MD 20742 \\
$^{3}$ Max-Planck-Institut fur Gravitationsphysik,
Albert-Einstein-Institut, Golm, Germany 
}
\ead{channa@phys.lsu.edu, megevand@phys.lsu.edu, evano@umd.edu, carlos@lsu.edu}
\begin{abstract}

Recent advances in the description of compact binary systems
have produced gravitational waveforms
that include inspiral, merger and ring-down phases. Comparing
results from numerical simulations with those of post-Newtonian (PN), 
and related, expansions
has provided motivation for employing PN waveforms in near merger epochs
when searching for gravitational waves and has encouraged the development
of analytic fits to full numerical waveforms.  The models and simulations
do not yet cover the full binary coalescence parameter space.  For these
yet un-simulated regions, 
data analysts can still conduct separate inspiral, merger and
ring-down searches.  
Improved knowledge about the end of the inspiral phase, the beginning of
the merger, and the ring-down frequencies could increase the efficiency of
both coherent inspiral-merger-ring-down (IMR) searches and searches over each
phase separately.
Insight can be gained for all three cases 
through a recently presented theoretical calculation, 
which, corroborated by the numerical results, 
provides an implicit formula for the final spin of the
merged black holes, accurate to within 10\% over a large parameter space.
Knowledge of the final spin allows one to predict the end of the inspiral
phase and the quasinormal mode ring-down frequencies, and
in turn provides information about the bandwidth and duration of the
merger.  In this work we will discuss a few of the implications of this
calculation for data analysis.

\end{abstract}
\maketitle

\section{Introduction}
Gravitational radiation emitted during the 
inspiral, merger and ring-down 
of two compact objects with a total mass below 
$\sim 100 M_{\odot}$ is a likely detection source
for interferometric gravitational wave detectors \cite{LastThree}.  
Maximal signal-to-noise ratio (SNR) is achieved in searches for compact binary
coalescence with matched filtering banks of template waveforms 
\cite{Owen:1996,CBC:S1, CBC:S2, CBC:S2BBH, CBC:S3S4} allowing faint signals
to be detected reliably.  
The SNR depends on having 
a template waveform that matches the true signal correctly \cite{Owen:1996}.
Because templates are weighted by the noise 
power spectral density of the detector with non-trivial
frequency dependence it is important {\it when} in the binary evolution the
template matches well.   
Numerical simulations
are improving our knowledge of the full waveforms emitted during the
inspiral, merger and ring-down phases of binary black hole coalescence \cite{IMR1,IMR2,Vaishnev}, yet
they still do not cover an adequate parameter space for constructing filter
banks \cite{Baker,Vaishnev}.  When the full parameter space is simulated
it will be useful to have fully parameterized models of these waveforms and
even now many have considered hybrid waveforms that patch together PN
and numerical simulations \cite{panetal,ajith,IMR1,IMR2,Vaishnev} as  
data analysis templates. Others have introduced a
phenomenological fourth order correction to make effective one-body (EOB)
waveforms match  numerical relativity results near the merger 
\cite{panetal,IMR1} known as {\it pseudo} 4PN (p4PN).

Before the full parameter space is 
simulated it is still possible to handle the three binary
coalescence epochs - inspiral,
merger and ring-down - separately.  We will show that a recent calculation
\cite{BKL07} 
which has power to predict the final spin of a binary merger, can
provide both insight into when the inspiral phase ends and what the 
quasi-normal ring-down frequency of the merger product should be for a large,
yet un-simulated, parameter space.  We will discuss the following
consequences of this calculation:
\begin{itemize}
\item{  
The calculation of the final spin given in \cite{BKL07} contains 
an implicit reference to
the innermost stable circular orbit (ISCO) 
of a test particle orbiting the Kerr
black hole that results from the  merger.  
We will 
also discuss the impact of using the new ISCO frequency as a cut
off for current gravitational wave searches that are targeting 
coalescing compact objects including the impact on SNR and the ability to
probe neutron star properties.}

\item{Knowing the final spin gives an estimate for the quasi-normal mode
ring-down frequencies.  These frequencies could be used in a separate matched filter search \cite{Jolien}.}

\item{
We can also use the final spin to calculate the light ring (smallest,
unstable circular orbit for a photon orbiting a Kerr black hole) and
quasinormal mode ring-down frequencies of the final black hole; we will see
that the l=2, m=2, n=0 QNM is quite close to the light ring frequency. We
will define a merger epoch as the evolution from ISCO to light ring and
dicuss how to obtain its time frequency content for yet unsimulated
systems.

We define the merger epoch as everything that occurs between the ISCO andi
ring-down frequencies.  We show that the duration of the radiation seems to be well described by the infall duration of a test particle orbitting the merged black hole just below the ISCO.  Knowing the duration and bandwith gives a time-frequency volume, which is useful in data analysis for unmodeled signals.}
\end{itemize}

\section{Formalism}
\label{s:formalism}
The inspiral phase of two compact objects is generally considered to end 
when the binary has evolved
to an innermost stable circular orbit (ISCO), should it exist.  
The ISCO can be defined
by waveform models and numerical relativity \cite{KWW,Blanchet,IMR1}.
Unfortunately there is a difference in the numbers obtained from
different methods. 
Some LIGO searches for non-spinning systems 
have taken the conservative approach to use the
ISCO defined for a test particle orbiting a Schwarzschild black hole. 

Intuitively the space-time of an inspiralling comparable mass binary 
system will not be well
described by a Schwarzschild space-time, and thus the Schwarzschild ISCO
will not capture the dynamics correctly.  If the merger product is 
a black hole, it will be a Kerr-like black hole with a final spin
that is a function of the masses and spins of the components.  
Recently
\cite{BKL07} has proposed a simple set of assumptions that predict the final
spin of the black hole merger product to well 
within 10\% of numerical simulations based on first principle arguments. 
Implicitly in the derivation of \cite{BKL07} is the ISCO radius of a 
test particle orbiting a Kerr black hole having a final spin equal to the 
spin of the merger product.  
The fact that the ISCO radius used gives the correct answer for the final 
space-time suggests that it plays an important role in the pre-merger dynamics.
We hypothesize that the test particle ISCO of the merged Kerr black
hole may describe a way to define the
end of the inspiral phase for binaries that produce black holes. 

The ISCO 
solution for a test particle orbiting a Kerr black hole is \cite{BPT72},
\begin{eqnarray}
\label{eq:KerrISCO}
   Z_1 &\equiv& 1 + \left(1 - \frac{a_f^2}{M^2}\right)^{1/3} 
   \left[(1+ \frac{a_f}{M})^{1/3} + (1- \frac{a_f}{M})^{1/3}\right]
\nonumber \\
   Z_2 &\equiv& \left(3 \frac{a_f^2}{M^2} + Z_1^2\right)^{1/2} \nonumber \\ 
   r_{\rm ISCO} &=& M\{ 3 + Z_2 \mp [(3-Z_1)(3+Z_1 + 2Z_2)]^{1/2} \}~~,
\end{eqnarray}
where $M$ is the total mass of the black hole and $a_f$ is the angular
momentum.  
The final spin is required in order to compute this ISCO. 
In \cite{BKL07}, assuming the amount of mass and
angular momentum radiated beyond the ISCO is small, the following
implicit formula 
for the final angular momentum of a black hole $a_f$ with component
spins aligned with the orbit is calculated,
\begin{equation}\label{eq_af}
   \frac{a_f}{M} = \frac{L_{orb}}{M^2} \left(q,\frac{r}{M}=\frac{r_{\rm ISCO}}{M},\frac{a_f}{M}\right)
                    + \frac{q^2\chi_1+\chi_2}{(1+q)^2}~~,
\end{equation}
where $\chi_i = a_i/m_i$, $q = m_1/m_2 \in [0,1]$ and $M=m_1 + m_2$ is the 
total mass. The implicitly found $a_f$ agrees well with numerical 
simulations \cite{BKL07} and the analysis
can be modified to include arbitrary spin angles. 
$L_{orb}$ is the orbital angular momentum contribution calculated
from the orbital angular momentum of a particle at the ISCO of a Kerr black 
hole with spin parameter $a_f$, which has the following 
expression \cite{BPT72},
\begin{equation}
  \frac{L_{orb}}{M^2}\left(q,\frac{r}{M},\frac{a_f}{M}\right) = \frac{q}{(1+q)^2}\frac{\pm(r^2 \mp 2 a_f M^{1/2} r^{1/2} + a_f^2)}
  {M^{1/2} r^{3/4} (r^{3/2} - 3 M r^{1/2} \pm 2 a_f M^{1/2})^{1/2}}~~,
\end{equation} 
where the upper/lower signs correspond to prograde/retrograde orbits.
In order to agree with numerical simulations this function has to be evaluated at $r=r_{ISCO}$ given 
by equation (\ref{eq:KerrISCO}).  
Alternative radii in the above prescription -- 
like the photon-ring radius or 
Schwarzschild radius -- give a
prediction that is off the numerically obtained value.
This fact,
combined with the same analysis extended to elliptic binaries, again
depending sensitively on the ISCO location \cite{SperhakeEtAl}, 
suggests the ISCO radius determined by this approach is relevant\cite{luis}.

In order to get the gravitational wave frequency at ISCO we use the coordinate angular velocity of a circular orbit \cite{BPT72},
\begin{equation}\label{eq:omega}
   \Omega = \pm \frac{ M^{1/2}} {r^{3/2} \pm a_f M^{1/2}}~~,
\end{equation} 
with $a_f$
given by the implicit equation (\ref{eq_af}).  The gravitational wave frequency
at a given radius is then $f_{\rm} = \Omega/\pi$, 
and so we define $f_{\rm ISCO \, [BKL]}$
as the frequency obtained by solving the system of equations (\ref{eq:KerrISCO}-\ref{eq:omega}) at the Kerr ISCO radius.

\begin{figure}[h!]
\begin{center}
\includegraphics[angle=-90,width=0.95\textwidth]{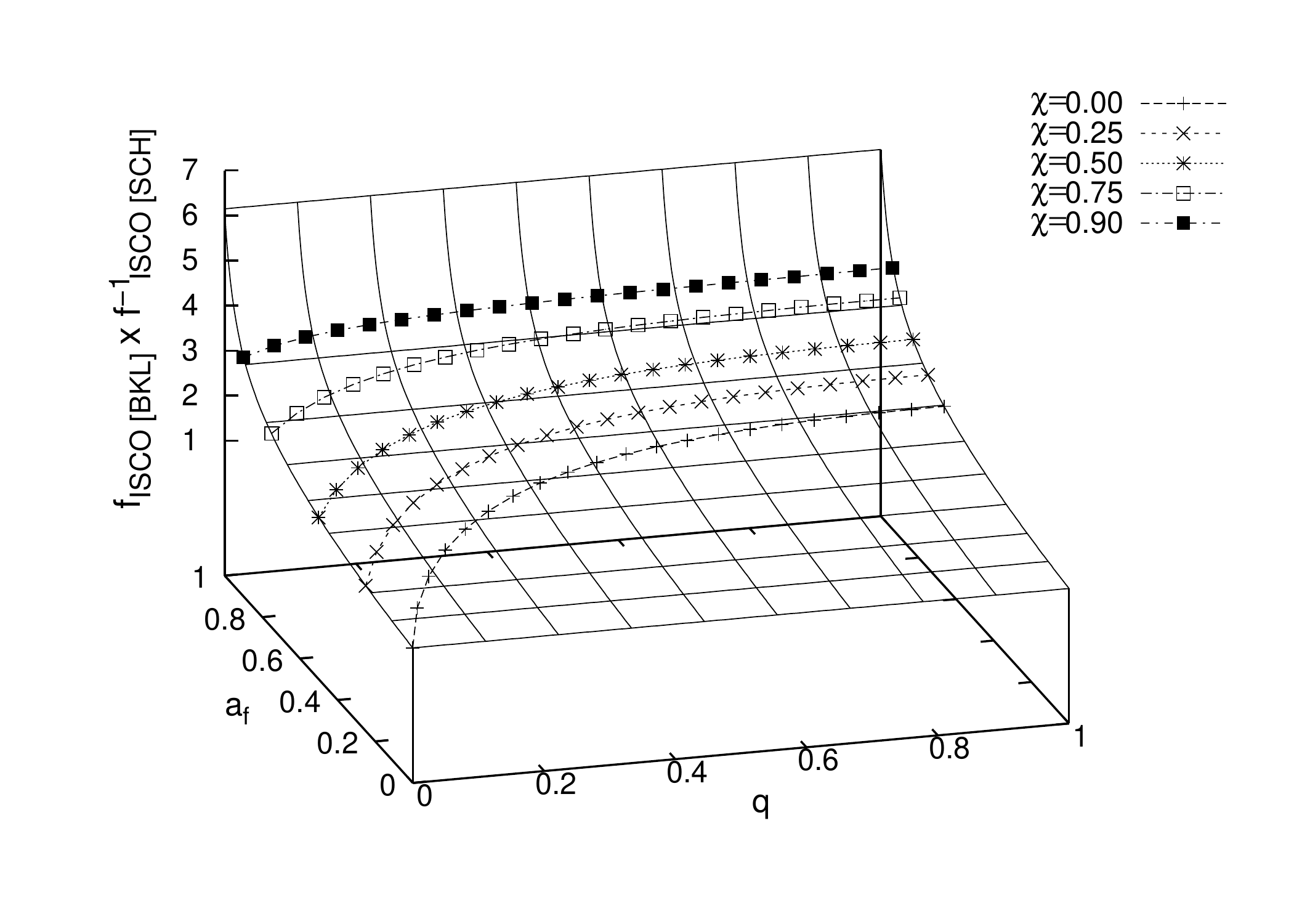}
\caption{\label{f:BKLsurface} The surface of solutions of the frequencies at the
ISCO as a function of the mass ratio $q$ and the final spin $a_f$ for components
with spins that are aligned with the orbital angular momentum. Also shown, are
curves corresponding to the solution of the equal spin case $\chi_1=\chi_2=\chi$.}
\end{center}
\end{figure}

The solution space for $f_{\rm ISCO \, [BKL]}$ can be written as a function of the final unknown spin $a_f$.
For convenience, we prefer to extend it as a surface parametrized by
$(a_f,q,\chi)$, as it is shown 
in figure \ref{f:BKLsurface}. This way the mass ratio dependence
of the lines corresponding to different individual spins
can be seen explicitly.  
For the equal mass case without spin the approximate expression for the Kerr
test particle ISCO frequency is,
\begin{equation}
\label{eq:BKLISCO}
f_{\rm ISCO \, [BKL]}(M,q) \approx 
\left(0.8 q^{3} - 2.6 q^{2} + 2.8 q + 1\right) \times \frac{1}{\pi (6 M)^{3/2}}~~,
\end{equation}
which can be compared to the Schwarzschild test particle ISCO 
$f_{\rm ISCO\,[SCH]} = \pi^{-1}(6M)^{-3/2}$.

Once the final spin is known it is possible to compute the last stable 
photon orbit (``light ring") and the quasi-normal mode ring-down frequencies 
\cite{Faithful}.  The light ring for a Kerr black hole is,
 \cite{BPT72}
\begin{eqnarray}
r_{\rm light} &=& 2M \{1+\cos\,[2/3\, \cos^{-1}\,(\mp a_f / M )]\} \\
\Omega_{\rm light} &=& \pm \frac{ M^{1/2}} {r_{\rm light}^{3/2} \pm a_f M^{1/2}}~~,
\end{eqnarray}
An approximate fit to the quasinormal mode ring-down frequencies as a
function of the final spin is given in \cite{Echeverria,Jolien,Berti}. 
For the l=2, m=2, n=0 mode we have,
\begin{equation}
\label{eq:qnm}
\Omega_{\rm QNM} \approx \frac{0.5}{M}\left(1-0.63(1-a_f)^{0.3}\right)~~,
\end{equation} 
Others ring-down modes could could be computed as necessary.  

\begin{figure}[h!]
\begin{center}
\includegraphics[width=0.95\textwidth]{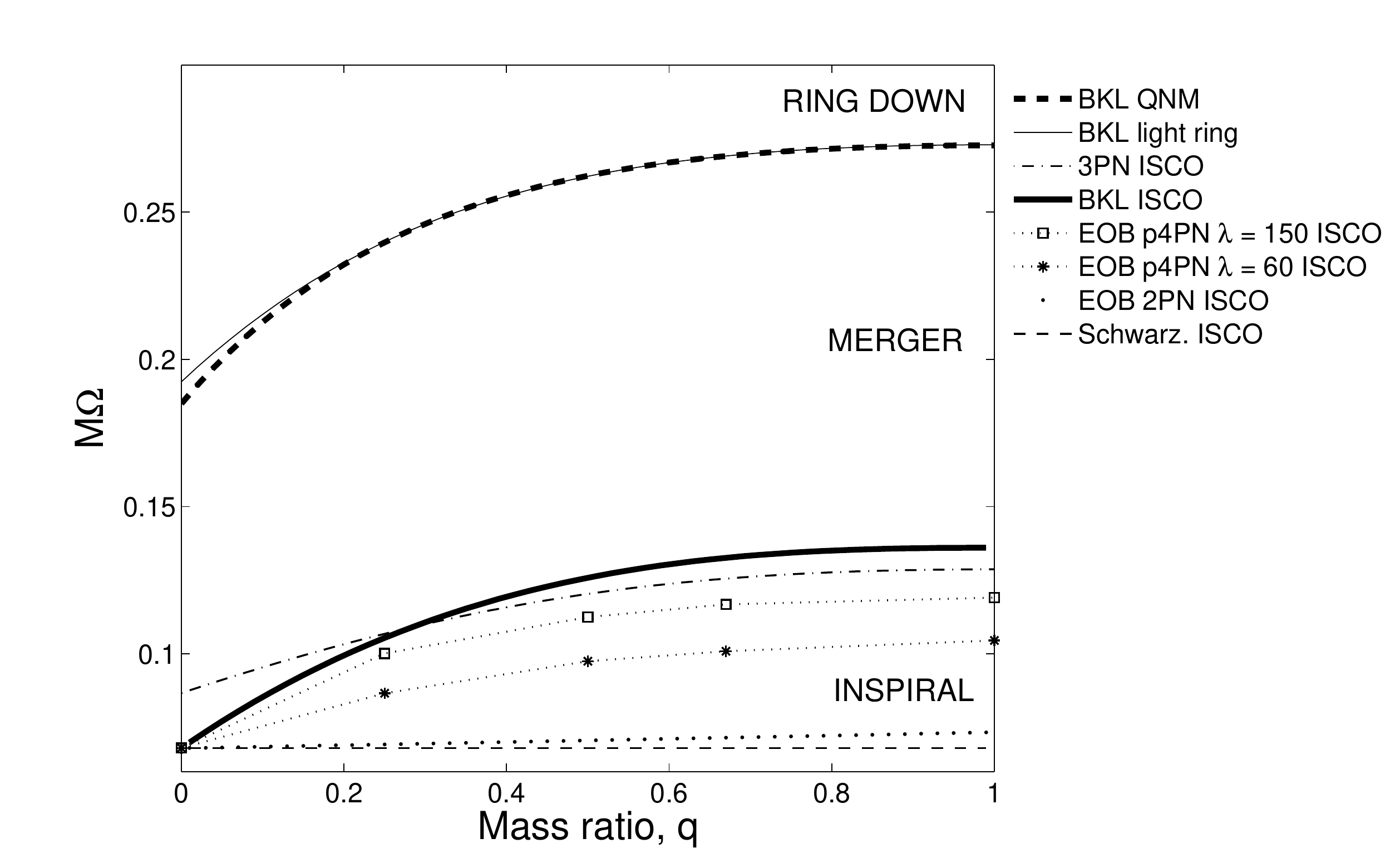}
\caption{\label{pn_isco}ISCO and QNM frequency estimates for 
non-spinning binaries as a 
function of mass ratio using different methods.  Knowing the final spin
of the black hole gives the expected ring-down frequency which agrees
with the light ring frequency.  The ISCO frequencies are very different
depending on the method.  The PN minimum
energy condition gives an inconsistent result in the extreme mass ratio
limit whereas the the other methods (EOB, BKL) agree with the 
Schwarzschild ISCO at small values of q.  The ISCO and QNM frequencies
define a natural merger epoch which can be analyzed, albeit sub-optimally,
even without knowing the numerical waveform.  } 
\end{center}
\end{figure}

Figure \ref{pn_isco} compares the BKL ISCO for non-spinning binaries with
the ISCOs computed from the minimum energy conditions at 3PN order
\cite{LRBlanchet} as well as 2PN and p4PN EOB ISCOs \cite{EOB,Faithful}.
BKL and EOB 
agree exactly with the Schwarzschild ISCO at the extreme mass ratio limit, 
but the PN calculations are known
to be inconsistent \cite{LRBlanchet} in that regime. Predicting the final
spin gives the $l=m=2$ ring-down mode frequency by (\ref{eq:qnm})
and the light ring frequency. Both are shown in figure \ref{pn_isco}.
The ISCO and light ring set a natural ``merger'' epoch which can be searched for by burst search techniques
\cite{S4Burst}
{\it in between} the inspiral and ring-down matched filter searches.
The ring-down frequencies shown in \cite{Faithful} agree well with these
predictions.  Although figure \ref{pn_isco} is plotted for the non-spinning
case the formalism described above can be easily generalized.   
 
It has been shown \cite{ExcessPower, interplay,FlanaganHughes} 
to be useful to understand un-modeled time-frequency
evolutions in gravitational wave analysis in terms of time-frequency
volumes, $\Delta f\,\Delta t$.  This concept
can provide insight into the merger epochs of various systems.  We will
consider three timescales for the merger
1) the Newtonian free fall time scale $t_{\rm ff}$,
2) the quadrupole radiation time scale $t_{\rm quad}$ and 
3) the time scale of a test particle following its geodesic just
inside the ISCO of the 
Kerr space-time (corresponding to the merger product). 
For comparison we also plot the time-frequency volumes found from 
the figures in \cite{Faithful}.
The Newtonian free fall time-frequency volume is,
\begin{equation}
\Delta f_{\rm ISCO} \,\Delta t_{\rm ff} 
= \frac{2^{-3/2}}{\sqrt{M\Omega_{\rm ISCO}}}~~.
\end{equation}
The quadrupole time-frequency volume is calculated by
assuming that an ISCO doesn't exist and that the system continues
to be driven by radiation.  To calculate this we use the quadrupole 
approximation to the inspiral waveform given by \cite{FinnChernoff}, which
gives the frequency evolution as,
\begin{eqnarray}
M\Omega = \left[\frac{5(1+q)^2}{256\,q}\frac{M}{t-t_o}\right]^{3/8}
\end{eqnarray}
This leads to the time frequency volume,
\begin{eqnarray}
\Delta f_{\rm BKL\,ISCO} \,\Delta t_{\rm quad} 
&=& \frac{5\,(1+q)^2}{256\,\pi\,q} 
\left[{(M\Omega_{\rm ISCO})^{-8/3}} - {(M\Omega_{\rm light})^{-8/3}}\right] \nonumber \\
& &
\times \left[(M\Omega_{\rm light}) - (M\Omega_{\rm ISCO}) \right]~~,
\end{eqnarray}
where the ISCO frequencies and light ring 
frequencies are obtained from the
BKL approximation as in figure \ref{pn_isco}.

\begin{figure}[h!]
\begin{center}
\includegraphics[width=0.95\textwidth]{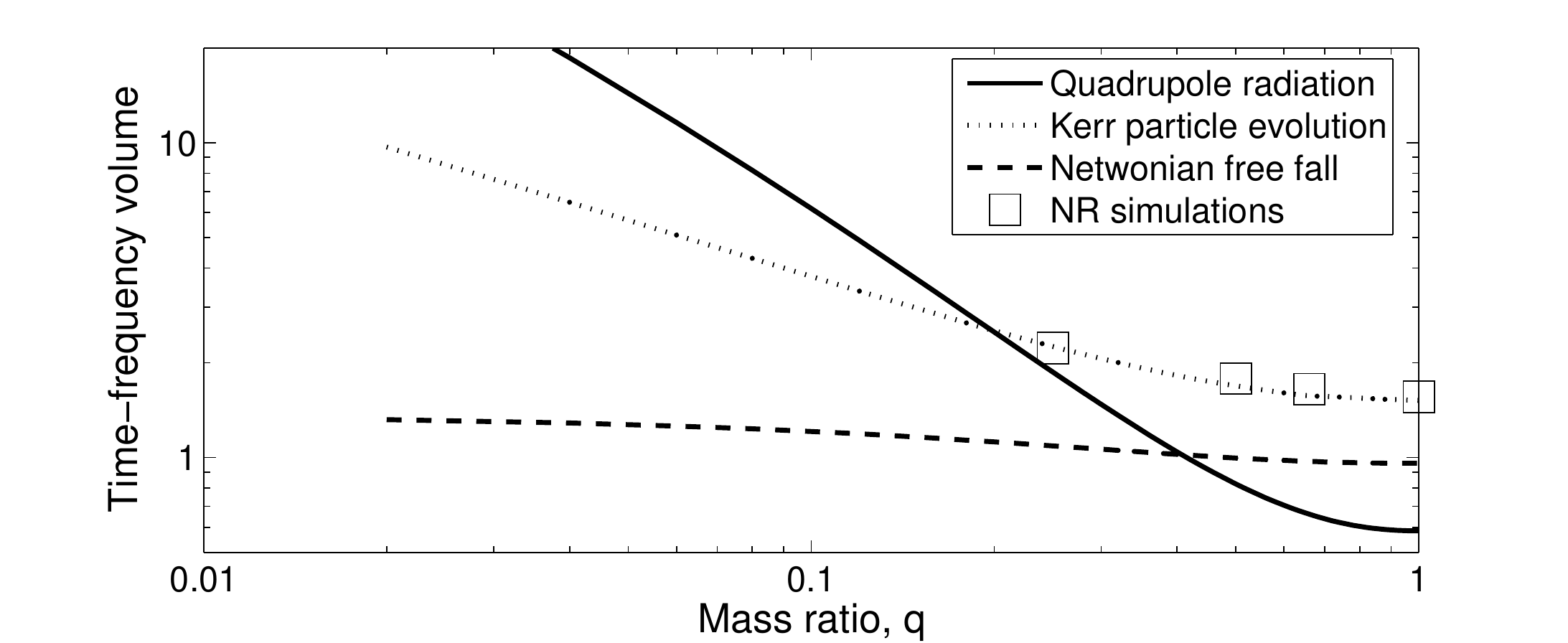}
\caption{\label{TFVolume}An estimate of the time-frequency volume for the 
merger epoch of non-spinning binary black holes.  
Numerical relativity results agree well with the time-frequency volume
of a test particle falling into the merged Kerr black hole from a circular orbit just inside the ISCO.
}
\end{center}
\end{figure}

The merger
evolution will not be purely dominated by radiation 
by virtue that there are no 
circular orbits below the ISCO. The numbers obtained
by this estimate should greatly over predict the time scale for
extreme mass ratios where particles would take an asymptotically
infinite time to fall in. 
The free fall time scale will under predict the time since
the system may still complete some (unstable) orbits before merging.  
The estimate that agrees best with numerical relativity simulations is
the infall time of a test particle falling from
a circular orbit just below the ISCO to the light ring of the
the merged Kerr black hole.  
Figure \ref{TFVolume} shows the
time-frequency volume for the merger epochs as a function of mass ratio
for non-spinning binaries.  The figure shows that 
the numbers taken from the simulations presented
in \cite{Faithful} agree well with the predictions from the 
test particle in the Kerr space-time.  
By these estimates the mergers from yet un-simulated evolutions (low mass ratio)
are likely to have the largest time frequency content and searches
may benefit from the numerical simulations of these \cite{interplay}. 
This information should be useful
in conducting IMR searches and also in guiding the construction of
analytic full waveform models. It is of course possible to repeat this analysis
for arbitrary spin configurations using the above prescription and others
\cite{BKL07,spin,RezzollaEtAl,Boyle}. 
We will leave further discussion of these matters to future work and 
will spend the rest of this work considering 
the implications of using the ISCO found from the above set of equations as a 
termination frequency for PN waveforms in initial LIGO data analysis.


\section{Possible impact on data analysis for the inspiral phase of compact
binary coalescence searches}
\label{s:data_analysis}


As mentioned previously, 
there is ambiguity on how to define when the inspiral phase ends.  
However, what is really necessary
is the characteristic frequency at which a given template waveform 
ceases to resemble the numerical simulations. 
Some searches for low mass binary systems of non-spinning component
masses
use the ISCO of a test particle orbiting a Schwarzschild black hole
\cite{FINDCHIRP}.  
Other searches use the ISCOs calculated from explicit PN energy considerations,
and some abandon the use of an ISCO altogether and use the Schwarzschild light ring as a termination frequency \cite{EOB}. 

Recall that for low mass ratio ($q\sim 0$) systems of non-spinning objects, 
the Schwarzschild test particle limit is a good approximation for the 
expected ISCO frequency 
since the merger product will be a Schwarzschild-like black hole.  
However, for systems with comparable masses and/or spin the true 
ISCO frequency may be different since the non-trivial contribution from the
orbital angular momentum will have a strong impact on the final black hole's
spin, and the space-time in the near merger epoch.    
Many have
addressed how well various PN approximants stay faithful to numerical
relativity solutions\cite{panetal, PN-NR, bakeretal, boyleetal}.  
Most approximants stay
faithful through the Schwarzschild ISCO frequency \cite{boyleetal}.
Some approximants fare extremely well beyond this point \cite{boyleetal}.  

The fact that some approximants do remain faithful far beyond the 
Schwarzschild ISCO leads us to examine using the ISCO frequency 
described in the previous section, equation (\ref{eq:BKLISCO}), 
as the termination frequency for inspiral data analysis.  As shown, this
frequency is consistent with some of the PN and EOB models for predicting 
the ISCO in the equal mass regime (fig. \ref{pn_isco}) 
and is consistent with exact solutions in 
the test particle limit.  It also has the advantage of being waveform model,
or fit, independent (based on first principles) and is easy to calculate.
We will conservatively model the impact of possible phase errors 
incurred by using an approximant that doesn't quite match numerical relativity results in our assessment.

\begin{figure}[h!]
\begin{center}
\includegraphics[width=0.95\textwidth]{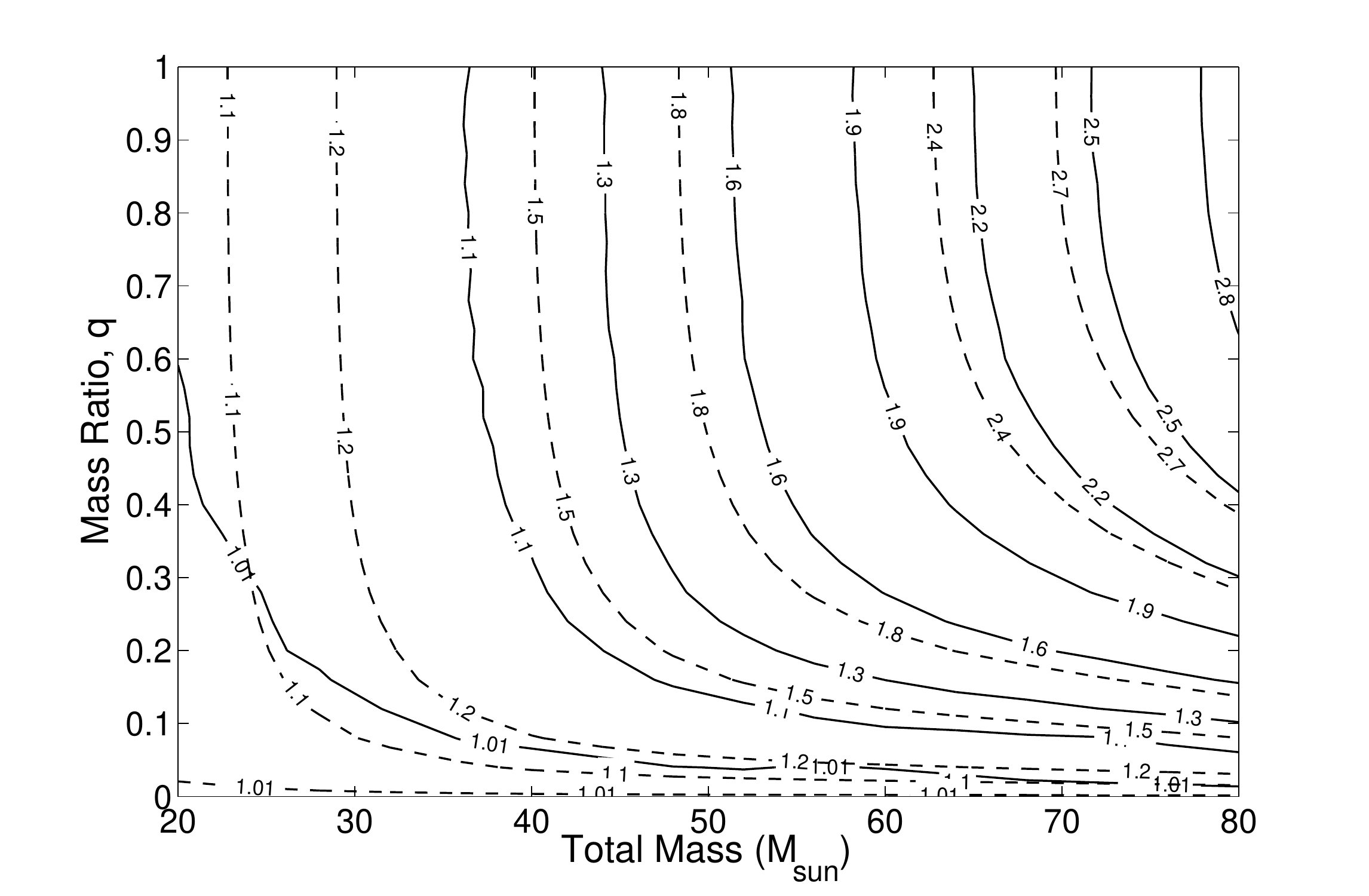}
\caption{\label{f:SNRratioMvsQ} The SNR ratio defined by equation (\ref{eq:SNRratio}) for the 
non-spinning case $\chi=0$.  The solid lines include phase errors of the
form (\ref{eq:SNRratio}) whereas the dashed lines assume perfect phase 
coherence for comparison.}
\end{center}
\end{figure}

We use the
stationary phase approximation waveforms, which are often employed in searches, 
and can be defined as,
\begin{eqnarray}
\label{eq:BNSwaveform}
\tilde{h}(f) & \propto & 
f^{-7/6} e^{i \Psi(f;M,\mu)}~~,
\end{eqnarray}
where $M$ is the total mass and $\mu$ is the reduced mass 
\cite{FINDCHIRP}. In order to ascertain the phase fidelity we turn to
comparisons given in \cite{panetal} which show for equal mass binary
simulations the phase difference between numerical results and 3.5 PN waveforms
as a function of frequency.  The figures in \cite{panetal} 
indicate a good fidelity in phase
through the Schwarzschild ISCO frequency.  However substantial
phase errors may accumulate between the Schwarzschild and
BKL ISCO frequencies.  In order to accurately predict the impact of integrating
PN waveforms to the BKL ISCO we decided to model the possibility 
that secular 
phase errors could be $\pm\pi$ or more radians between
the Schwarzschild and BKL ISCOs.
We propose calculating
the SNR ratio of an ensemble of waveforms where phase errors are allowed
to accumulate linearly between the Schwarzschild ISCO $f_2(M,q)$ and the 
BKL ISCO $f_1(M,q)$ given by,
\begin{equation}\label{eq:SNRratio}
\frac{{\rm SNR}_{f_1}}{{\rm SNR}_{f_2}} = \left<
\left(\int_{f_{\rm low}}^{f_1} \frac{e^{\pi i R_n (f_1-f_2)/f_1 \,\Theta(f_1-f_2)}\,{\rm d}f}{f^{7/3}\,S_n(f)}\right)^{1/2}
\left(\int_{f_{\rm low}}^{f_2} \frac{{\rm d}f}{f^{7/3}\,S_n(f)}\right)^{-1/2}
\right>~~,
\end{equation}
where $R_n$ is a normally distributed variable with mean 0 and variance 1.  
$f_{low}$ is 40Hz \cite{CBC:S3S4}.
\footnote{We take $S_n(f)$ to be approximated analytically as \cite{Damour},
\begin{equation}
S_n(f) \approx 
9\times 10^{-46}\, \left[ \left(\frac{4.49 f}{150}\right)^{-56} 
                       + 0.16 \left(\frac{f}{150}\right)^{-4.52} 
                       + 0.32 \left(\frac{f}{150}\right)^2 + 0.52 \right]~~.
\end{equation}
}
We note that the above calculation is a conservative estimate of the SNR
gain for two reasons:  1) A template with different parameters may match
the numerical waveforms better, producing better phase agreement.  Since 
search results are maximized over SNR this is an important possibility.  
2) The phase errors do not accumulate linearly between these frequencies, most
of the phase error occurs near the BKL ISCO. 
Using the \{$M,q$\} 
dependent ISCO frequencies as $f_1(M,q)$ and the Schwarzschild
ISCO frequencies as $f_2(M)$ we plot the expected SNR ratio for initial LIGO
\cite{SRD} in  
figure \ref{f:SNRratioMvsQ}, which shows that for some total mass and mass ratio
combinations there is an appreciable gain in SNR by integrating to the new
ISCO frequencies despite modeled phase errors.  
This calculation is not intended to suggest that SPA templates cut at the
BKL ISCO produce the optimal gain in SNR, since other approximants or
frequency cut offs could do better.
Instead, it aims to simply illustrate the effect of integrating
to the BKL ISCO as a function of $M$ and $q$ for present searches.  
The mass range is limited to 80 $M_{\odot}$ so that at least two cycles 
exist in the waveform between 40Hz and the Schwarzschild ISCO frequency.  


\section{Probing the tidal disruption of neutron stars}
\label{s:disruption}
If the BKL ISCO truly marks the transition from an inspiral, radiation-dominated evolution to a dynamical one then it does have some consquence
for the ability to probe the tidal disruption of neutron stars through
gravitational waves.
Estimation of gravitational-wave frequencies for tidal disruption of 
NS-BH binaries may be useful for determining various properties of the 
neutron star, such as the radius \cite{vallisneri,saijo-nakamura}. 
Knowing the radius of 
the neutron star would in turn provide information about its equation of 
state \cite{lindblom,harada2001}. However, 
as pointed out in \cite{vallisneri}, it would be difficult to
extract information  about the disruption 
unless it occurs before the binary reaches the ISCO. 
If the disruption occurs afterwards, its signature in the
produced gravitational waves might be too weak for extracting accurate
information. Thus, the ``useful'' cases are those for which the tidal-disruption frequency is less than the
frequency at the plunge ($f_{\rm td}<f_{\rm plunge}\equiv f_{\rm ISCO}$). 

A more accurate calculation of $f_{\rm plunge}$ would imply a better estimate of the range of NS-BH binaries for which 
the disruption is more plausible to be measured.
In this section we show how that range 
changes  when the formalism of section~\ref{s:formalism} is applied, as compared to that in \cite{vallisneri}.

\begin{figure}[h!]
\begin{center}
\includegraphics[angle=-90,width=0.95\textwidth]{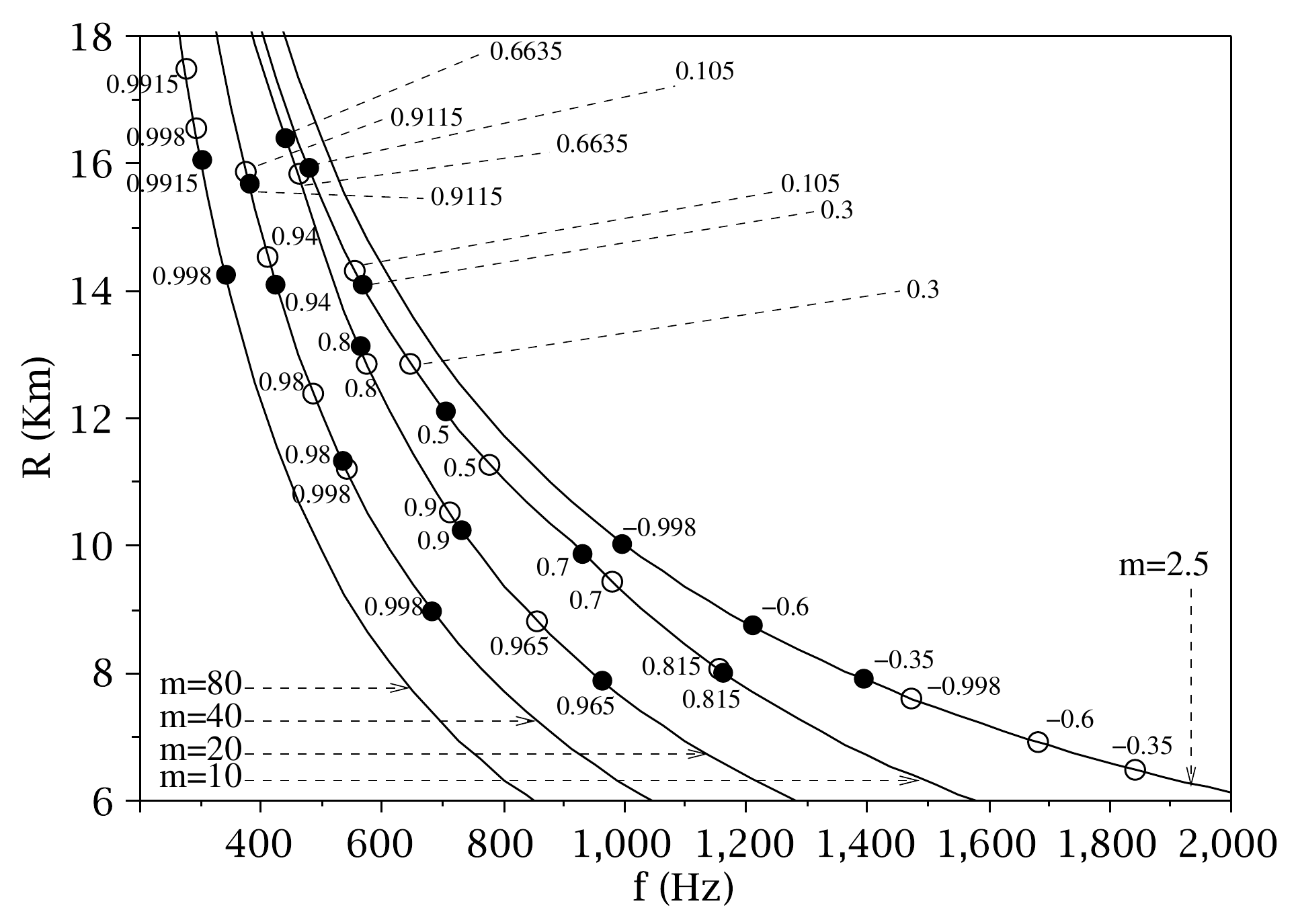}
\caption{\label{f:Rvsf} Radius of the neutron star vs. the disruption frequency. The circles indicate the points at which the 
plunge would occur; the filled circles correspond to $f_{\rm plunge}$ as calculated in \cite{vallisneri}, included here for
comparing. The value of $\chi_{\rm BH}$ is indicated next to the corresponding circle where possible, 
otherwise a dashed line was used to connect the value to the circle. Each curve is labeled with the value of 
the BH mass: $m_{\rm BH}=2.5-80 {\rm M_\odot}$.  }
\end{center}
\end{figure}

The frequencies obtained in \cite{vallisneri} correspond to the extreme mass ratio case, $q=0$, and 
non-spinning neutron star, $\chi_{\rm NS}=0$. In figure~\ref{f:Rvsf} we reproduce those results (compare to figure~2 of \cite{vallisneri}),
together with the values obtained using the formalism presented in this work. We also assume in this case that the neutron 
star is non-spinning, but with $q=m_{\rm NS}/m_{\rm BH}\neq0$. The curves of $R$~vs~$f_{\rm td}$ are calculated 
using the formula given in equation.~(3) of \cite{vallisneri}, for the same parameters as in that work: $m_{\rm NS}=1.4M_{\odot}; m_{\rm BH}=2.5, 10, 20, 40, 80M_{\odot}$. The circles over each curve specify where the plunge 
occurs for different values of $\chi_{\rm BH}$. Filled circles correspond to the extreme mass ratio approximation. 
The meaningful part of each curve will be that to the left of the white circles ($f_{\rm td}<f_{\rm plunge}$).

The plunge frequencies calculated in \cite{vallisneri} for the case with $m_{BH}=2.5M_{\odot}$ (black circles in the 
right-hand-side curve of figure~\ref{f:Rvsf}) correspond to retrograde orbits. 
However, for that mass ratio ($q=0.56$), the test particle orbiting the merger
product ISCO does so in a prograde fashion due to the orbital angular momentum 
dominating the final spin. 
Thus, the final spin is aligned with the orbital angular momentum even though its initial spin is anti-aligned.

We now compare with the NR results obtained in \cite{ShibataTaniguchi}. We calculate 
$f_{\rm BKL}$ for the six cases considered 
in in that work, models A-F. The BKL frequencies are shown in table~\ref{t:shibata}, 
together with the corresponding parameters $q$ and $M \equiv M_{\rm BH}+M_{\rm NS}$ 
($\chi_{\rm BH}$ and 
$\chi_{\rm NS}$ are set to zero as in \cite{ShibataTaniguchi}).
Although the models studied in \cite{ShibataTaniguchi}
expand only a small range of parameters;
we see that,
at least in that range, the frequencies obtained with 
the BKL approach are consistent with the NR results (See figure~7 of 
\cite{ShibataTaniguchi}) - that is they predict in every case a higher 
value than the measured $f_{\rm cut}$ caused by the tidal effects.

\begin{table}[h!]
\caption{\label{t:shibata} BKL frequencies for the same parameters as models A-F of 
\cite{ShibataTaniguchi}. These values are consistent with the NR results - 
that is the BKL ISCO frequency is higher than the observable $f_{\rm cut}$
from the simulations \cite{ShibataTaniguchi}}
\centering
\begin{tabular}{c c c c c}
\hline \hline
Model & $q$ & $M(M_\odot)$ & $f_{\rm BKL}({\rm kHz})$ & $f_{\rm cut}$ from \cite{ShibataTaniguchi}  \\
\hline
A & 0.327 & 5.277 & 1.38 & 1.16 \\
B & 0.327 & 5.244 & 1.39 & 1.41 \\
C & 0.328 & 5.311 & 1.37 & 0.92 \\
D & 0.392 & 4.623 & 1.65 & 1.14 \\
E & 0.392 & 4.594 & 1.66 & 1.40 \\
F & 0.281 & 5.929 & 1.18 & 1.09 \\
\hline \hline
\end{tabular}
\end{table}


\section{Conclusion}
\label{s:conclusion}
The estimate for the spin of the black hole that results from 
BH-BH, BH-NS, and some NS-NS mergers found in \cite{BKL07} leads to an
estimate for the ISCO, light ring, and quasinormal mode ring-down frequencies
of a general compact binary system that is waveform
model independent.  Both frequencies 
have impact on searches for the inspiral, merger and ring-down 
epochs of the gravitational waves emitted by these systems.  We have
shown that the most interesting merger-epoch time-frequency volumes 
are for extreme mass ratios.  More work may have to be done to model merger
epochs for these systems.  The formalism presented here to describe
inspiral, merger and ring-down epochs can be extended to arbitrary spin
configurations and may help guide the construction of better analytic models.

As for inspiral epoch searches alone, we have shown that for symmetric mass,
non-spinning systems with greater than 45 $M_{\odot}$ total mass there can
be a 30\% gain in SNR over using the Schwarzschild ISCO radius
as is done currently \cite{FINDCHIRP} even including conservative
phase errors.  
It is worth noticing that more general
fittings can be computed in order to cover the full range of parameters
$(q,\chi_1,\chi_2)$ within the BKL approximation.
Additionally we discussed cases where neutron star tidal disruption may be observed more easily through gravitational waves. We would like to stress that 
this work is open ended.  Here we have only 
considered spins that are aligned and anti-aligned with the orbital
angular momentum.  It is certainly possible to extend this work to make
predictions about more general cases, all of which should be compared with
numerical relativity results as they are produced.    

\section{Acknowledgements}
The authors would like to acknowledge L. Lehner for suggesting this project and for invaluable discussions.  The authors also thank
L. Lehner and F. Pretorius for
providing the test particle time of flight used in section 2.  
Gabriela Gonz\'alez, Patrick Brady, Alessandra Buonanno and Jolien Creighton
provided motivating discussions and insightful comments.
C. Hanna would like to thank the 
LIGO Scientific Collaboration Compact Binary 
Coalescence (LIGO CBC) working group. This work was
supported in part by NSF grants PHY-0605496, PHY-0653369 and PHY-0653375 to 
Louisiana State University and PHY-0603762 to the University of Maryland.
C. Hanna would like to thank the Kavli Institute
for Theoretical Physics, for their hospitality, where some of this work was
completed.  The Kavli Institute is supported by NSF grant PHY05-51164.

\end{document}